\documentclass[referee,useAMS,usenatbib]{mn2e}
\usepackage{graphicx}

\newcommand{\nitrogen}{[N~{\sc ii}]}

\title{3--D modelling of the collimated bipolar outflows of compact planetary nebulae 
with WR--type central stars} 

\author[S. Akras et al.]
{S. Akras$^{1}\thanks{e-mail:akras@astrosen.unam.mx}$ and  
J. A. L\'{o}pez$^{1}$\\
$^{1}$Instituto de Astronomia, Universidad Nacional Autonoma de
Mexico,Apdo Postal 877, Ensenada 22800, Baja California, Mexico}

\begin{document}  

\date{Received **insert**; Accepted **insert**}

\pagerange{\pageref{firstpage}--\pageref{lastpage}}

\maketitle
\label{firstpage}

\begin{abstract} 

We present high--resolution, long--slit spectroscopic observations of five 
compact ($\leq$ 10 arcsec) planetary nebulae located close to the galactic bulge 
region and for which no high spatial resolution images are available. The data 
have been drawn from the San Pedro M\'artir kinematic catalogue of galactic planetary 
nebulae (L\'opez et al. 2012). The central star in four of these objects ( M 1--32, 
M 2--20, M 2--31 and M 3--15) is of WR--type and the fifth object (M 2--42) has a 
wels type nucleus. These observations reveal the presence in all of them of a 
dense and thick equatorial torus-like component and  high--speed, collimated, 
bipolar outflows. The code SHAPE is used to investigate the main 
morpho--kinematic characteristics and reproduce the 3--D structure of these objects
assuming a cylindrical velocity field for the bipolar outflows 
and a homologous expansion law for the torus/ring component. The deprojected expansion 
velocities of the bipolar outflows are found to be in the range of 65 to 200 km 
$\rm{s^{-1}}$, whereas the torus/ring component shows much slower expansion velocities, 
in the range of 15 to 25 km $\rm{s^{-1}}$. It is found that these planetary nebulae 
have very similar structural components and the differences in their emission line 
spectra derive mostly from their different projections on the sky. The relation of 
their morpho--kinematic characteristics with the WR--type nuclei deserves 
further investigation.

\noindent
\end{abstract}


\section{Introduction}

Low to intermediate mass stars (0.8 to 8 $\rm{M_\odot}$) undergo spectacular 
structural changes during the last phases of their evolution. 
 According to the interacting stellar wind model 
(ISW; Kwok Purton and Fitzgerald 1978), the spherically symmetric PNe
are formed by the interaction of two isotropic stellar winds, 
a slow and dense one from the Asymptotic Giant Branch (AGB) phase and a fast 
and tenuous one during the PN phase.  The generalized ISW model considers in addition 
the contribution of an equatorial density enhancement at the exit of the AGB phase that 
produces a density contrast leading to the formation of axisymmetric shapes (e.g. Balick 
1987) that may range from mildly elliptical to bipolar. In fact, the majority of 
planetary nebulae (PNe) and proto--PNe (PPNe) show axisymmetric morphologies. In some 
cases highly collimated, high--speed bipolar outflows are also found. The causes of the 
equatorial density enhancement and the jet-like outflows are still under 
debate (e.g. Balick \& Frank 2002) and the two most likely being the presence of magnetic 
fields (e.g. Garcia--Segura \& L\'opez 2000, Frank \& Blackman 2004) and post-common 
envelope, close binary nuclei (e.g. Soker \& Livio 1994, De Marco 2009).
Sahai and Trauger (1998) proposed as a shaping mechanism for the bipolar 
and multi--polar PNe, the presence of highly collimated outflows developed during the 
post--AGB or PPNe phase. All these elements represent the main considerations in recent 
morphological classification studies of PNe (e.g. Parker et al. 2006, Miszalski et al. 
2008, Sahai et al. 2011, Lagadec et al. 2011). However, imaging alone can be in some cases 
deceiving in describing the real shape of a PN due to the inherent uncertainty introduced 
by the effects of the projection on the plane of the sky for a  three dimensional nebula. 
The simplest example is that of an axisymmetric nebula, such as a bipolar, with a thick 
waist observed pole-on, in which case the nebula appears as a round doughnut projected on the sky.
In these cases spatially resolved, high spectral resolution spectroscopy becomes an ideal 
tool to explore the three dimensional structure of the nebula by examining the Doppler 
shifts in the emission line profile and assuming in a first approximation a homologous 
expansion for the nebula. Most of these morpho-kinematic studies have been performed on 
relatively  large, spatially resolved PNe (e.g. L\'opez et al. 2012, Clark et al. 2010, 
Garc\'{\i}a--D\'{\i}az et al. 2009) but they can also be very revealing when studying spatially 
unresolved, compact PNe, as we show here.

\begin{table*}
\centering
\caption[]{Source  Properties}
\label{table5}
\begin{tabular}{llllllllllll}
\hline 
Name    & PNG & RA & Dec & distance (kpc)$\rm ^a$ & V$_{\rm{sys}}($km $\rm{s^{-1}}$) & central star$\rm ^b$ \\   
\hline
M 1--32 & 011.9+04.2  & 17 56 20.1 & -16 29 04.6 & 4.8$\pm$1.0 &  $-105$ &  [WO4]pec \\
M 2--20 & 000.4--01.9 & 17 54 25.4 & -29 36 08.2 & 9.5$\pm$1.9 &  $+60$& [WC5-6]     \\
M 2--31 & 006.0--03.6 & 18 13 16.1 & -25 30 05.3 & 6.3$\pm$1.3 &  $+ 155$&  [WC4]     \\
M 2--42 & 008.2--04.8 & 17 01 06.2 & -34 49 38.6 & 9.5$\pm$1.9 &  $+ 105$& wels      \\
M 3--15 & 006.8+04.1  & 17 45 31.7 & -20 58 01.8 & 6.8$\pm$1.4 &  $+ 100$& [WC4 ]    \\
\hline
\end{tabular}
\medskip{}
\begin{flushleft}        
${\rm ^a}$ Stanghellini \& Haywood 2010\\
${\rm ^b}$ Acker \& Neiner 2003
\end{flushleft}
\end{table*}

In this work, we perform a morpho--kinematic study of five, relatively bright, compact PNe 
with no discernable structure and with seeing limited angular sizes ranging from 5 to 10 
arcsec. No high spatial resolution images for these objects were found in the literature or 
the usual repositories of images of PNe. These objects were chosen from the the San Pedro 
Martir kinematic catalogue of Galactic Planetary Nebulae (L\'{o}pez et al. 2012) on the basis 
of their line emission spectra that show the presence of fast, collimated bipolar outflows. 
The objects selected are: M 1--32, M 2--20, M 2--31 and M 2--42 and M 3--15.

Based on their galactic coordinates, distances and systemic velocities they seem located in 
the Galactic bulge or close to it, see Table 1. The central stars for four of them have been 
classified as Wolf-Rayet type (Acker \& Neiner 2003) and the fifth one as a weak emission--line 
star or wels (Tylenda, Acker \& Stenholm 1993). As mentioned above, the long-slit, spectroscopic 
observations reveal the presence of highly collimated, fast, bipolar outflows surrounded by a 
thick equatorial enhancement, as a torus or a ring. We combine these data with the 3--D 
morpho--kinematic code SHAPE (Steffen \& Lopez 2006, Steffen et al. 2011) to analyze  the 
3--D structure of these outflows and the relation of their appearance with different projection 
on the sky. 

In Section 2, the observation and data reduction are presented. In 
Section 3, we describe the parameters used in the morpho--kinematic code SHAPE as 
well as the modelling results. We finish by summing up the results of this work 
in Section 4.

\section{Observations}

High--resolution, long-slit spectra of the PNe M 1--32, M 2--20, M 2--31, M2--42 and M 3--15
were obtained using the Manchester echelle spectrometer, MES--SPM (Meaburn et al. 2003)
on the 2.1 m telescope at the Observatorio Astron\'omico Nacional at San Pedro Martir 
Observatory in Baja California, Mexico, in its $f$/7.5 configuration.  The observing runs 
for M 1--32, M2--42, M 2--20,  M 2--31 and M 3--15 took place in 2009 July, 2007 June, 2006 
July, 2004 June and 2004 June, respectively. In all cases MES-SPM was equipped
with a SITE--3 CCD detector with 1024$\times$1024 square pixels, each
24 $\mu$m on a side ($\equiv$0.312 arcsec pixel$^{-1}$).  A 90
\AA~ bandwidth filter was used to isolate the 87$^{th}$ order containing the
H$\alpha$ and \nitrogen{} $\lambda$$\lambda ~$6548,~6584~\AA, nebular
emission lines. Two-by-two binning was employed in both the
spatial and spectral directions. Consequently, 512 increments, each
0\farcs624{} long gave a projected slit length of 5\farcm32 on the
sky. The slit was 150 $\mu$m{} wide ($\equiv$~11~km $\rm{s^{-1}}$ and
1\farcs9). The slit was oriented north -- south (P.A. $= 0\degr$) for M 1--32, 
M 2--20 and 3--15. For M 2--31 and M 2--42 two slit positions are available, 
P.A. $= 35\degr$ and P.A. $= 0\degr$ and P.A. = $48\degr$ and P.A. = $-45\degr$, respectively.  
All the spectroscopic integrations were of 1800~s duration. The wavelength calibration was 
performed using a Th/Ar calibration lamp to an accuracy of $\pm$ 1 km $\rm{s^{-1}}$  when converted 
to radial velocity. The stellar continua were not subtracted. Only in two cases, M 2--42 and M 2--31, 
a very faint continuum could be discerned within the very short spectra range of our single--order 
echelle spectra. The data reduction was performed using the standard IRAF routines. Individual 
images were bias subtracted and cosmic rays removed. All spectra were also calibrated to heliocentric 
velocity. These spectra are part of the SPM Kinematic Catalogue of Galactic Planetary Nebulae; L\'{o}pez 
et al. 2012) and are available at http://kincatpn.astrosen.unam.mx where images from the digitized sky 
survey are also shown with the location of the slits overlaid for each object.

\section{SHAPE modelling}

Modelling with SHAPE follows three main steps. First, defining the geometrical forms to use;
SHAPE has a variety of objects such as a sphere, torus, cone, cube, etc. whose basic forms can be 
modified by the user (e.g. squeeze, twist, boolean, etc). Second, an emissivity distribution is 
assigned to each individual object or structure, and third, a velocity law is chosen as a 
function of position. SHAPE gives as result a two dimensional image and synthetic P--V arrays, 
which are rendered from the 3D model to be compared with the observed data. The parameters of 
the model are then iteratively adjusted until a satisfactory solution is obtained. 
Figure 1 presents the expected shapes of bi--dimensional line profiles or position--velocity 
diagrams (PV) from a spatially resolved, long--slit, spectrum for some representative cases,
such as an expanding bubble, a face--on torus, a bipolar nebula with its lower part tilted into 
the plane of the sky and finally, the same bipolar with an equatorial torus.

Figure 2 presents the observed \nitrogen  ~ emission line spectra with the corresponding 
synthetic P--V diagrams. These bi--dimensional line profiles clearly show the structure of fast, 
bipolar outflows and a central thick waist, presumably an equatorial density enhancement in 
the form of a torus or ring. This information is used in combination with the code SHAPE to 
iteratively find a combined geometry and expanding velocity field solution that replicates 
the observed position -- velocity diagrams of the bi-dimensional line profiles. 
The resultant synthetic P--V diagrams are shown side by side to the observed ones, to their right, 
in the panels in Figure 2. These  were reproduced assuming an axisymmetric  geometry and a 
Hubble--type velocity law increasing linearly with distance from the central star to reconstruct 
the outflows along the polar direction. A key characteristic of cylindrically symmetric and 
homologously expanding nebulae is that the P--V diagrams can be stretched in velocity in such a 
way that the outline of the image and the P--V diagram match. This match determines the factor of 
proportionality in the mapping between position and velocity. It must however be noted that the SHAPE 
model cannot constrain the extent of the nebula along the sightline and thus there can be more than 
one solution to the fit. Moreover,  the information in the present cases comes from the line profiles 
alone (no morphology information) from single slit  cuts over the nebula and without a previous 
knowledge of the position angle of the main axis of the nebula. To our only advantage is the slit 
width of two arcsec that includes a significant portion of the core of these compact nebulae, 
providing sufficient information to investigate their inner structure. Considering these limitations 
we have therefore assumed reasonable sizes for all objects and considered the observed expansion 
velocities as lower limits. Under these conditions we have selected the simplest solution that best 
fit the data.

The equatorial thick torus/ring component was assumed to expand radially from the central star. 
Similar solutions have been used for NGC 6337 (Garc\'{i}a--D\'{i}az et al. 2009) and and more recently 
for BD +30 3639 (Akras \& Steffen 2012). 

In order to reproduce the line profiles shown in Figure 2, the inclination of the symmetry axes of 
M 1--32, M 2--20, M 2--31, M 2--42 and M 3--15 were found to be $5\degr$, $55\degr$, $5\degr$ , $77\degr$ 
and $5\degr$  respectively, with respect to the line  of sight, with an uncertainty in this case of 
the order of 20\%, beyond this figure the line profiles start to show significant tilt differences.

 In Figure 3, we present the SHAPE mesh models that best fit each of the PN under study, 
displaying the geometry of the system at four different orientations given by 
position angle and inclination: $0\degr, 0\degr; 0\degr, 60\degr; 60\degr, 60\degr\ \&\ 90\degr, 
90\degr$,  before rendering. All models share as common elements the bipolar outflows, the torus 
or thick ring and an outer shell.

 \section{Discussion on the individual objects}

The unusual, wide line profile of M 1--32 has been noted before by Medina et al. 2006. 
The P--V diagram shown in  Figure 2 (top row, left panels) for this object reveals 
fast collimated bipolar outflows reaching $\pm 200$ ~km $\rm{s^{-1}}$. The orientation of 
M 1--32 is nearly pole-on, $5\degr$ with respect to the line of sight as derived from the model, 
and the torus intersected by the slit is clearly revealed by the bright knots at the centre 
of the profile. The torus component is expanding radially with a velocity of $\approx$15 km 
$\rm{s^{-1}}$. Faint extensions beyond the bright equatorial knots disclose the presence of 
an outer, thin shell.

The P--V diagram for M 2--20 (Figure 2, top row, right panels) indicates a tilted axisymmetric 
structure, either an elliptical or a mild bipolar nebula. The central region is dominated by 
two bright equatorial knots  from a toroidal structure. The knots are displaced one respect 
to the other due to the inclination of the torus that shows the front and back sides approaching 
and receding due to its radial expansion. These knots have expansion velocities with respect 
to the core of $\approx$15 km $\rm{s^{-1}}$. The observed expansion velocity of the lobes with 
respect to the core is $\pm$ 65 km $\rm{s^{-1}}$ and for an inclination with respect to the line 
of sight of $55\degr$ the expansion velocity for the  bipolar outflow transforms into $\pm$ 
125 km $\rm{s^{-1}}$  and 25 km $\rm{s^{-1}}$ for the equatorial toroid.

M 2--31 shows for the slit oriented at P.A. = $35\degr$, a P--V diagram (Figure 2, middle row, 
left panels) dominated by two bright knots located at very nearly the same position along the 
slit and separated by 30 km $\rm{s^{-1}}$, fainter extensions to these knots reveal a collimated 
bipolar outflow reaching $\pm$ 90 km $\rm{s^{-1}}$. There is also a faint and seemingly, slightly 
tilted shell component that surrounds the bright knots or blobs. This tilt indicates an 
ellipsoidal shell expanding at a similar velocity to the bright knots. The slit oriented at 
P.A. = $0\degr$ (Figure 2, middle row, right panels) shows the same components except for the 
tilt in the faint shell which suggests that the slit at P.A. = $35\degr$ lies close to its major 
axis whereas the slit at  P.A. = $0\degr$ cuts across it. Contrary to the previous cases where 
the bright knots have been identified with emission from a thick toroidal structure, in this case 
the characteristics of the bi-dimensional line profile indicate that the bright knots are part of 
an expanding bipolar system, probably a pair of compact, bright lobes,  perhaps with a structure 
similar to the PN Mz 3 (or M 1--92) (Redman et al. 2000) but seen nearly pole-on and with material 
beyond the knots that expands at significantly larger speed.

For M 2--42 its P--V diagram reveals clearly a collimated bipolar structure with  both lobes 
showing knots or condensations at their tips (Figure 2, bottom row, left panels). The lobes have 
an extent of about 10 arcsec from the centre where again the presence of a  conspicuous equatorial 
torus is indicated by the line profile. The knots are observed to be expanding with a velocity 
of $\sim$$\pm$15 km $\rm{s^{-1}}$ with respect to the core. Considering the $77\degr$ angle of inclination 
derived from the model for this nebula, the polar knots expand at a deprojected velocity of 
$\pm$ 70 km $\rm{s^{-1}}$ from the core. The torus expands radially with a velocity of $\approx$15 km 
$\rm{s^{-1}}$, very similar to the case of M 1-32.

M 3--15 looks nearly as a twin brother to M 2--42 but observed close to the line of sight (Figure 2, 
bottom row, left panels). The bi-dimensional line profile in this case is dominated by a bright, 
compact, core containing a central toroid and a nearly pole-on, collimated outflow that extends to 
$\pm$ 100 km $\rm{s^{-1}}$. An inclination of $5\degr$ with respect to the line of sight is obtained 
here from the model.

\section{Conclusions}

In this work, we carried out a morpho--kinematic study of five compact PNe. The data were drawn 
from the Kinematic Catalogue of Galactic Planetary Nebulae (L\'opez et al. 2012). The sample 
was selected on the basis of compact PNe with no discernible or apparent structure from their 
ground--based images, no high spatial resolution imaging available (e.g. {\it HST}) and showing 
the presence of collimated, high-speed, bipolar outflows in their emission line spectra. 
Fortuitously, the five PNe selected under the criteria described above turned out to have WR--type 
central stars. It is worth pointing--out that the SPM catalogue lists at present 84 PNe 
with WR--type nuclei, 92 spatially resolved bipolar out of which 43 are labeled as showing 
fast projected outflows and 152 PNe from the galactic bugle region. It is our intention 
to conduct next a thorough search in the catalogue to select from the lists mentioned 
above additional PNe that match similar criteria adopted here.

In this work, we demonstrate that with practically no morphological information and limited 
spectral information consisting of long--slit, spatially resolved, echelle spectra combined 
with the reconstruction code SHAPE, a fair morpho-kinematic representation of the nebulae can 
be achieved. The process yields a first order description of the main morphological elements of 
the nebula, the inclination of the main outflow symmetry axis with respect to the line of sight 
and with it an estimate of the true expansion velocity for the bipolar outflow. We find that 
these planetary nebulae have very similar structural components and the differences in their 
emission line spectra derive mostly from their different projection on the sky and degree of 
collimation of the bipolar outflows. All cases studied here are characterized by the presence 
of an equatorial density enhancement or toroid, a collimated bipolar outflow and an outer shell. 
The bipolar expansion velocities range from 70 to 200 km $\rm{s^{-1}}$ whereas the tori are found 
to expand rather slowly, in the order of 15 to 25 km $\rm{s^{-1}}$.

The results of this work strongly indicate the importance of evaluating the true structure of 
compact PNe for statistical studies that consider the different morphological classes for evolution 
and population synthesis studies. Given the distance to the galactic bulge, there is a large 
population of unresolved PNe there that need to be investigated in detail. Whether the fact that 
all the compact PNe studied here turned out to have a WR-type nucleus has an statistical significance 
for the bulge PNe with collimated bipolar outflows or not deserves further investigation. 
Likewise a detailed reinvestigation of the kinematics of PN with WR--type nucleus such as the work 
of Medina et al. (2006) with spatially resolved, long slit spectra of higher spectral resolution 
should turn fruitful results and possibly establish a stronger link between PNe with WR--type 
nuclei and fast, collimated, bipolar outflows.
It is also interesting to note that Lopez et al. (2011) associate toroidal structures with close binary 
central stars and Miszalski (2011) also associates bipolar outflows with close binary nuclei in PNe.
Given the characteristics of the present sample, it would be interesting to search for the presence
of binary nuclei in these objects.
\\
\\
\\
We acknowledge the financial support of DGAPA-UNAM through grants IN110011 and IN100410. 
S. A. gratefully acknowledges a postdoctoral scholarship from DGAPA-UNAM.
The authors are grateful to the anonymous referee for his valuable comments 
and suggestions.

\bibliographystyle{mnras}

\begin{figure*}
\includegraphics[scale=0.65]{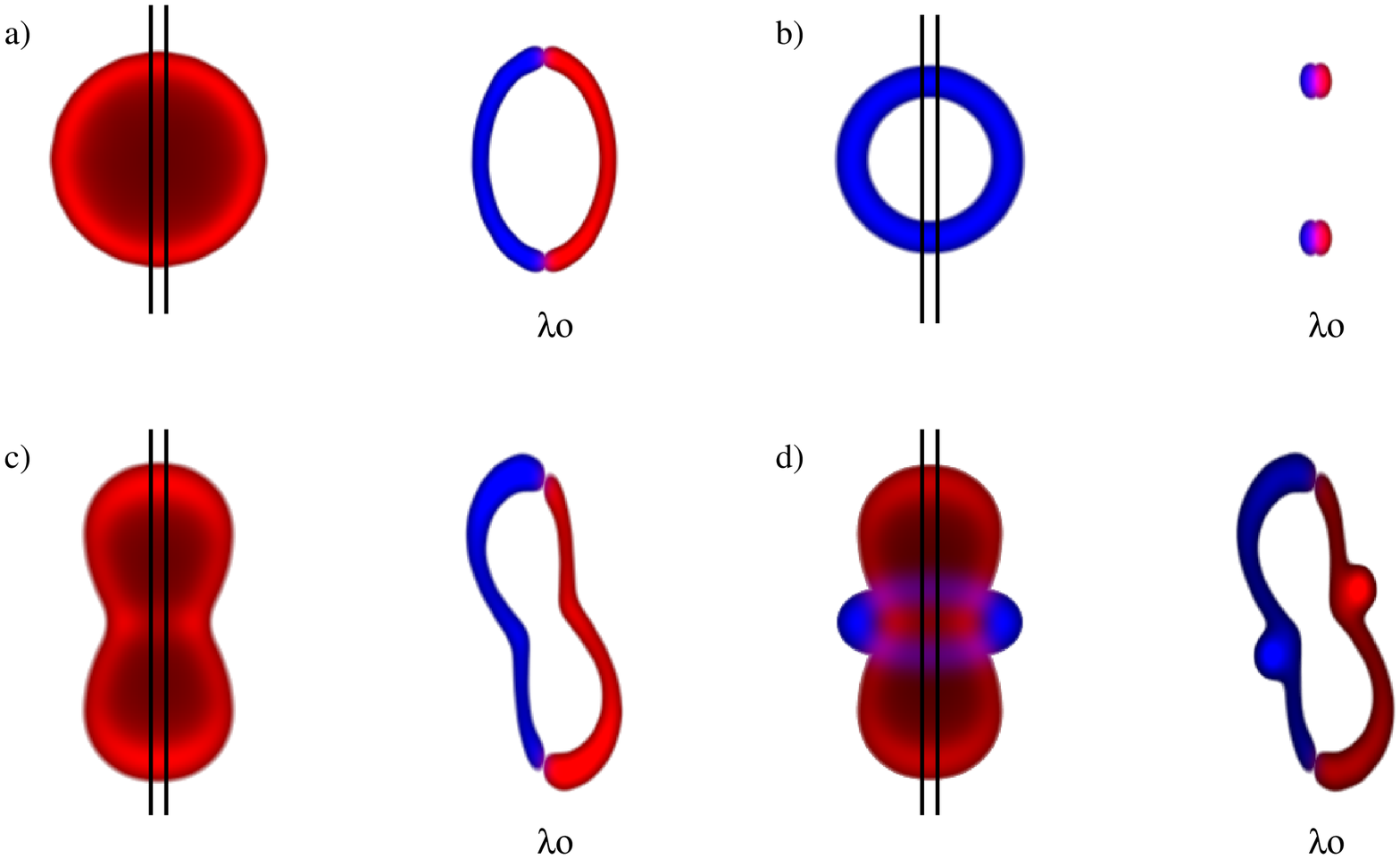}
\caption{Projected nebula images on the sky (left) and expected shape of
bi--dimensional line profiles or PV diagrams from
a spatially resolved, long--slit spectrum (right) for a) a spherically symmetric bubble,
b) a face--on torus, c) a sligthly tilted bipolar nebula and d) the same bipolar but with 
an equatorial torus. The position of the long--slit is indicated in each case.
(See the electronic version for the colour version of these figures).}
\end{figure*}

\begin{figure*}
\includegraphics[scale=0.42]{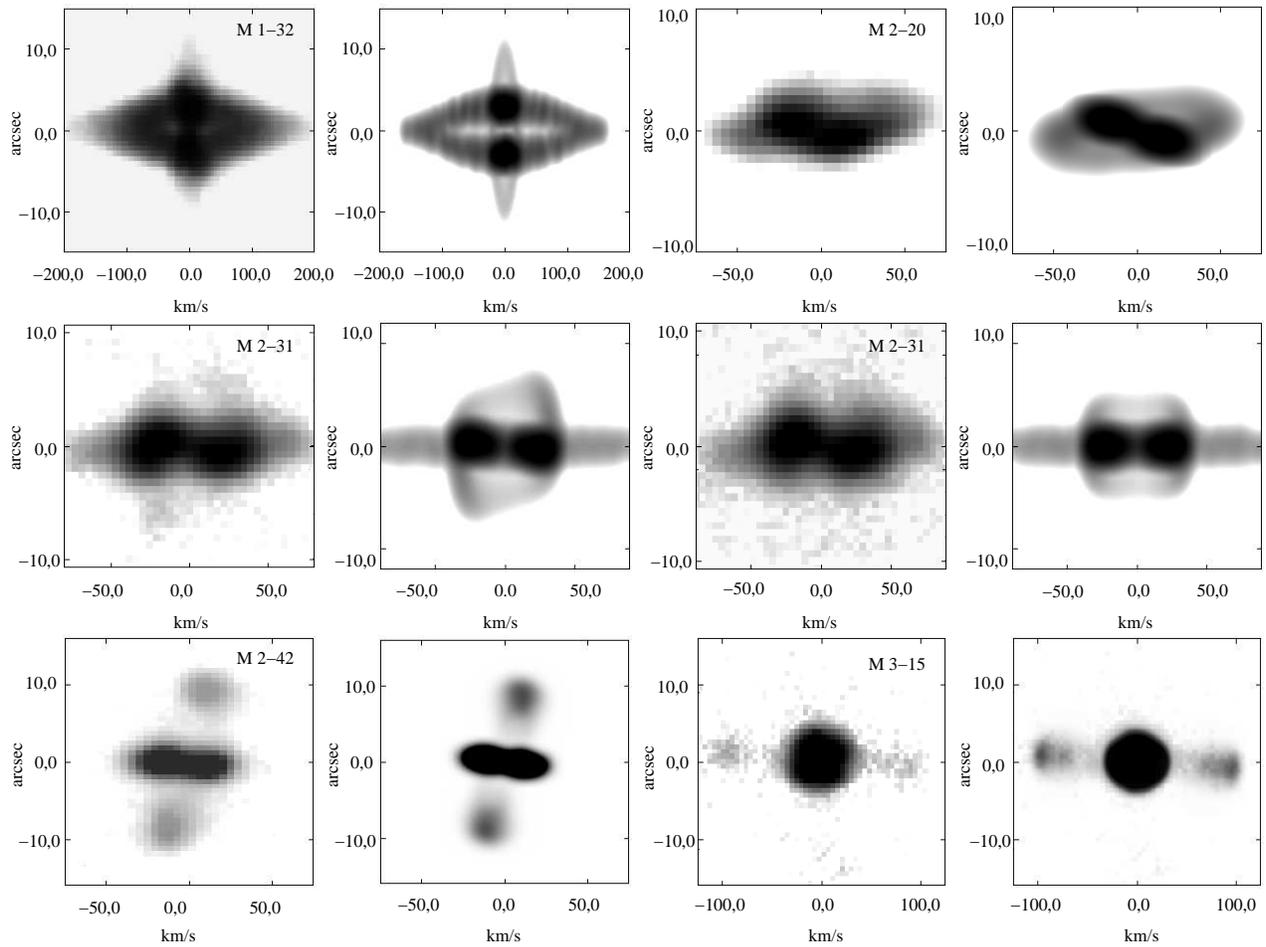}
\caption{The observed \nitrogen\ and corresponding synthetic P--V  diagrams of the planetary 
nebulae studied here are shown side by side. Top row : M 1--32 (P.A. = $0\degr$) and 
M 2--20 (slit P.A. = $0\degr$). Middle row: M 2--31 (P.A. = $35\degr$ and P.A. = $0\degr$). 
Bottom row: M 2--42 (slit P.A. = $48\degr$) and M 3--15 (P.A. = $0\degr$).}
\end{figure*}

\begin{figure*}
\includegraphics[scale=0.45]{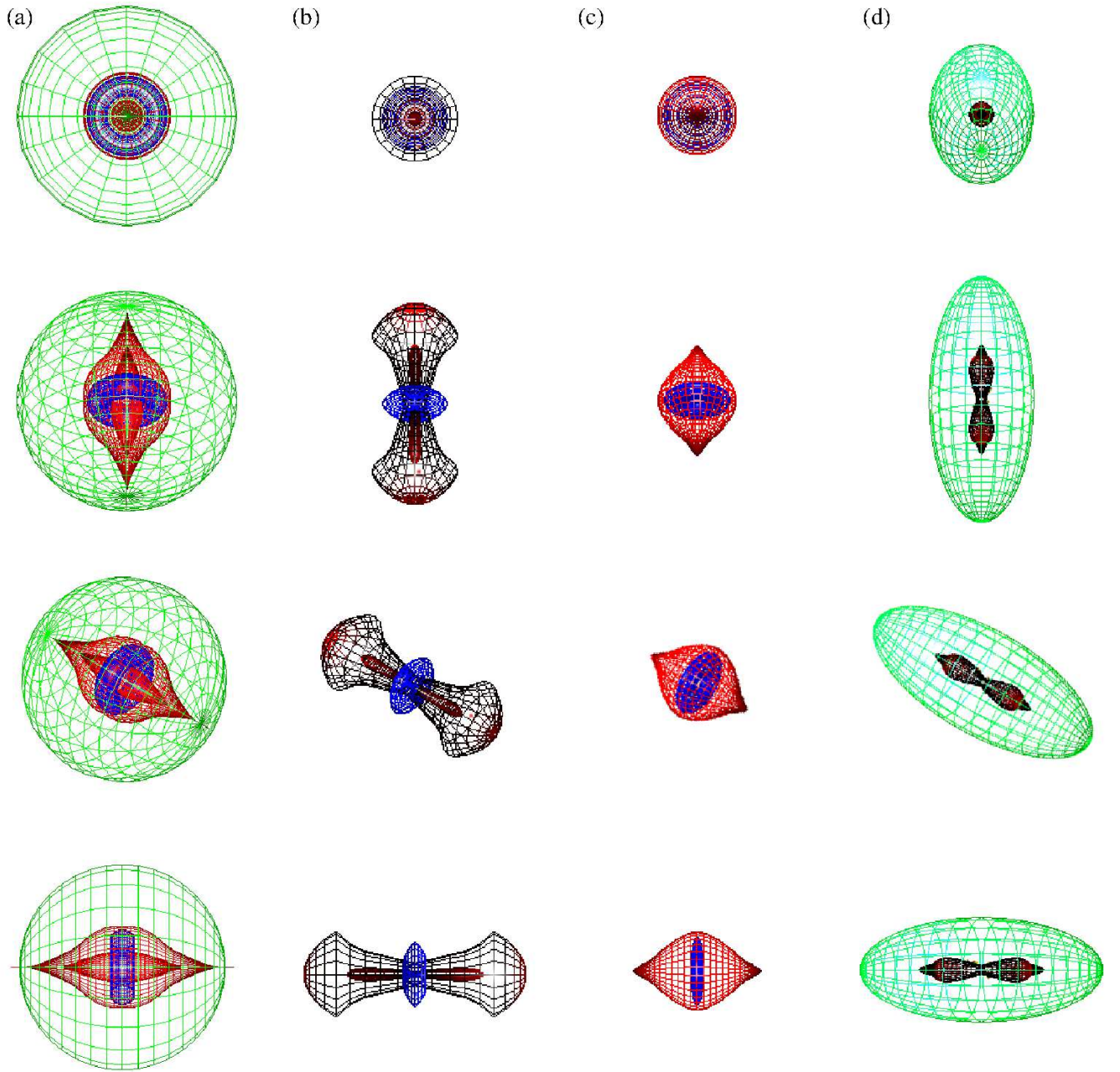}
\caption{SHAPE mesh models of M 1--32 (first column), M 2--42 and M 3--15 (second column),  
M 2--20 (third column) and M 2--31 (fourth column), 
before rendering at four orientations (position angle, inclination): 
0,0 (first row); 0,60 (second row); 60,60 (third row) and 90,90 (fourth row).
Each model consists (i) the outflows or the bipolar structure (red), (ii) the 
torus/ring component (blue) and (iii) the halo and outer shell (green). 
(See the electronic version for the colour version of these figures).}
\end{figure*}

\end{document}